\documentclass[conference]{IEEEtran}
\usepackage{hyperref}
\IEEEoverridecommandlockouts

\usepackage{graphicx}
\usepackage{booktabs}
\usepackage{cite}
\usepackage{amsmath}
\usepackage{algorithm}
\usepackage{algorithmic}
\usepackage{amssymb}
\usepackage{lipsum}
\usepackage{caption}
\usepackage{xcolor}
\usepackage{placeins}
\usepackage{booktabs} 
\usepackage{pifont}
\usepackage{multirow}
\usepackage{subfig}
\usepackage{hyperref}
\hypersetup{
    colorlinks = true,
    linkcolor = blue,
    urlcolor = blue,
}
\usepackage{enumitem}
\usepackage[T1]{fontenc}
\usepackage{CJKutf8}
\usepackage{url}
\usepackage{adjustbox}
\usepackage{cite}
\usepackage{amsmath,amssymb,amsfonts}
\usepackage{algorithmic}
\usepackage{graphicx}
\usepackage{textcomp}
\usepackage{xcolor}
\def\BibTeX{{\rm B\kern-.05em{\sc i\kern-.025em b}\kern-.08em
    T\kern-.1667em\lower.7ex\hbox{E}\kern-.125emX}}
\begin{document}

\title{Learning Spatial Awareness for Laparoscopic Surgery with AI Assisted Visual Feedback}

\author{\IEEEauthorblockN{1\textsuperscript{st} Songyang Liu}
\IEEEauthorblockA{\textit{Department of Civil \& Coastal Engineering} \\
\textit{University of Florida}\\
Gainesville, USA \\
liusongyang@ufl.edu}
\and
\IEEEauthorblockN{2\textsuperscript{nd} Yunpeng Tan}
\IEEEauthorblockA{
\textit{Farragut High School}\\
Knoxville, USA \\
yunpengbtan@gmail.com
}
\and
\IEEEauthorblockN{3\textsuperscript{rd} Shuai Li}
\IEEEauthorblockA{\textit{Department of Civil \& Coastal Engineering} \\
\textit{University of Florida}\\
Gainesville, USA \\
shuai.li@ufl.edu}

}

\maketitle

\begin{abstract}
Laparoscopic surgery constrains surgeons spatial awareness because procedures are performed through a monocular, two-dimensional (2D) endoscopic view. Conventional training methods using dry-lab models or recorded videos provide limited depth cues, often leading trainees to misjudge instrument position and perform ineffective or unsafe maneuvers. To address this limitation, we present an AI-assisted training framework developed in NVIDIA Isaac Sim that couples the standard 2D laparoscopic feed with synchronized three-dimensional (3D) visual feedback delivered through a mixed-reality (MR) interface. While trainees operate using the clinical 2D view, validated AI modules continuously localize surgical instruments and detect instrument-tissue interactions in the background. When spatial misjudgments are detected, 3D visual feedback are displayed to trainees, while preserving the original operative perspective. Our framework considers various surgical tasks including navigation, manipulation, transfer, cutting, and suturing. Visually similar 2D cases can be disambiguated through the added 3D context, improving depth perception, contact awareness, and tool orientation understanding.
\end{abstract}


\section{Introduction}
Minimally invasive surgery (MIS), particularly laparoscopic surgery, has become the standard in many clinical procedures due to its numerous benefits, including reduced patient trauma, faster recovery, and fewer postoperative complications~\cite{Mayo}. However, MIS presents a unique set of cognitive and perceptual challenges for surgeons. One of the most critical limitations is the lack of depth perception caused by reliance on monocular, single-axis 2D endoscopic video feeds~\cite{bogdanova2016depth}. This constraint significantly impairs spatial awareness, especially for novice surgeons, leading to errors such as grasping empty space, misaligned cutting, or unintended tissue clipping. These spatial misjudgments not only compromise surgical precision but also increase the risk of complications and prolong operative time.

Traditional 2D laparoscopic visualization imposes increased cognitive load and limits accurate spatial judgment. Without binocular depth cues, surgeons must mentally infer the 3D spatial relationships of tissues and instruments, resulting in frequent perceptual errors. Despite the proven benefits of stereoscopic 3D systems in improving spatial perception and reducing task times~\cite{yim2017three, beattie2021laparoscopic}, such systems are often limited to intraoperative use and are not commonly employed in surgical training environments.

Existing laparoscopic training typically relies on physical simulators, video-based observation, or cadaver-based practice. These methods face major limitations, including high cost, limited anatomical variability, and the absence of real-time spatial feedback. Virtual and augmented reality (VR/AR) platforms have emerged to address these challenges~\cite{hong2021simulation, rasheed2023low}, but most either immerse the trainee entirely in a virtual 3D environment or overlay annotations onto 2D videos without achieving real-time synchronization between the two views. Moreover, few systems actively detect spatial errors or provide actionable feedback during the training process.

To overcome these limitations, we propose an AI-assisted, dual-visualization surgical training framework that explicitly targets the development of 3D spatial awareness. The system combines standard 2D endoscopic video feeds with synchronized 3D anatomical visualizations rendered in NVIDIA Isaac Sim. Using a mixed-reality (MR) headset such as Meta Quest 3, trainees can simultaneously observe both views, bridging the perceptual gap between 2D and 3D domains. Within the Isaac Sim environment, surgical instruments and tissues are simulated with physically accurate dynamics, while real-time sensor and camera APIs continuously provide scene data for AI analysis and visual feedback generation. AI models perform real-time instrument detection, segmentation, and interaction recognition. Their outputs are integrated into Isaac Sim through the Action Graph system to generate dynamic visual feedback. Specifically, nodes such as \textit{On Contact Event}, \textit{Write Prim Material}, and \textit{Draw Debug Line} are triggered during tool-tissue interactions. These nodes visualize the end-effector trajectory, mark unsafe depths, and change surface colors to indicate correct or incorrect operations. For instance, a green overlay appears when the tool aligns correctly with the target, while a red highlight indicates excessive penetration or deviation. The \textit{Print Text} node displays real-time textual cues (e.g., “Correct operation” or “Unsafe depth”), reinforcing spatial reasoning through both visual and cognitive feedback. By integrating AI perception with Isaac Sim’s physics-based simulation, our system establishes a closed feedback loop for mixed-reality surgical training. Trainees receive immediate and intuitive feedback based on their actual performance, transforming traditional passive training into an interactive, perceptually rich learning experience. The framework effectively enhances depth awareness, spatial reasoning, and hand-eye coordination, while reducing dependence on instructor supervision. Using MR hardware and AI-driven perception, our work provides a scalable, automated platform for addressing the root causes of spatial errors in MIS.

\begin{figure}
    \centering
    \includegraphics[width=\linewidth]{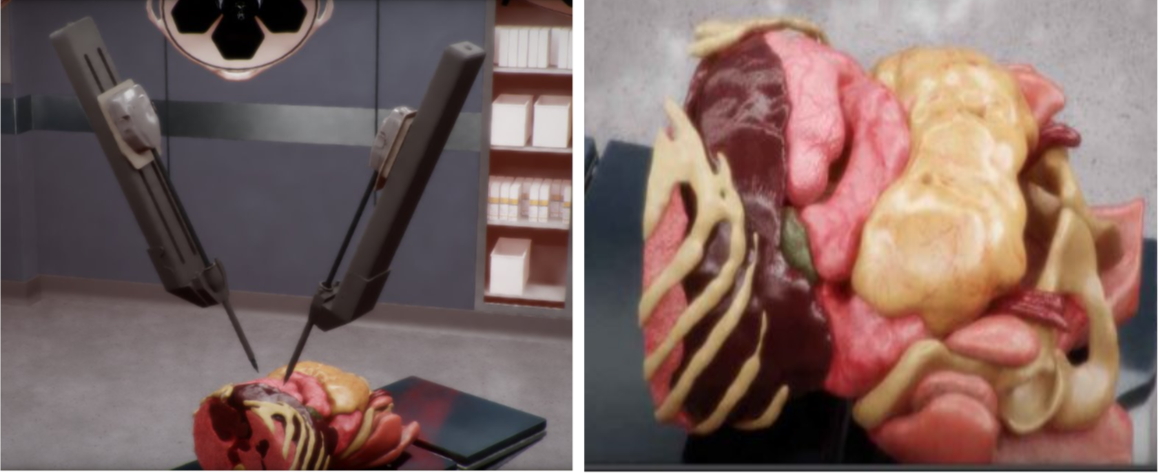}
    \caption{High-fidelity virtual surgical environment used in our simulation. The left panel shows a photorealistic surgical operation room with daVinci research kit Patient Side Manipulators (PSM), while the right panel presents detailed 3D models of human anatomical organs designed for realistic surgical training.}
    \label{fig:1}
\end{figure}

The remainder of this paper presents the design of the proposed framework, the integration of AI-based perception and feedback modules, and simulation-based validation in representative laparoscopic scenarios such as suturing and tissue manipulation. Our goal is to create a perceptually synchronized, feedback-rich training environment that enhances spatial awareness without requiring specialized imaging hardware or manual annotation. Fig.~\ref{fig:1} illustrates the photorealistic surgical room and anatomical models adapted from~\cite{moghani2025sufia}. The code for this work is available at~\href{here}{https://github.com/cgchrfchscyrh/ICIR\_2025}.

\section{Related work}
Traditional laparoscopic surgery presents a significant cognitive challenge for novice surgeons due to its reliance on monocular, single-axis 2D camera views. This restricted visualization inherently limits depth perception and spatial awareness, often leading to procedural errors such as ineffective grasping, inaccurate cutting, or inadvertent damage to surrounding structures. Beattie et al.~\cite{beattie2021laparoscopic} demonstrated that stereoscopic 3D visualization can significantly improve laparoscopic task performance and spatial accuracy among novices. Similarly, Yim et al.~\cite{yim2017three} showed that 3D laparoscopic systems reduce error rates and shorten procedure times by providing surgeons with improved depth cues. However, these systems primarily focus on intraoperative visualization upgrades and do not fundamentally address the perceptual training challenges present during early-stage skill development.

To address these training limitations, simulation-based surgical education has increasingly adopted VR and AR systems. Hong et al.~\cite{hong2021simulation} reviewed a wide range of laparoscopic simulators and recommended that next-generation training systems incorporate AR visualization and intelligent feedback mechanisms to enhance spatial understanding. Rasheed et al.~\cite{rasheed2023low} developed a Unity-based VR simulator with realistic 3D rendering and haptic interaction, which improved trainees' spatial comprehension in laparoscopic nephrectomy tasks. Colman et al.~\cite{colman2025lapar} demonstrated that an AR-enhanced box trainer providing virtual guidance overlays resulted in significant improvements in trainee performance and subjective spatial awareness.

Recent reviews have emphasized that many VR/AR-based systems either immerse trainees entirely in 3D environments or overlay information on 2D videos, often lacking real-time synchronization between 2D and 3D views~\cite{celdran2025use}. Moreover, most systems do not explicitly detect or correct spatial misjudgments during training. While AI-based systems appear in surgical education, such as Bogár et al.’s~\cite{bogar2024validation} low-fidelity VR trainer with automated skill assessment, they often assess performance post hoc and do not provide real-time spatial feedback.

In contrast, our approach explicitly targets the underlying cause of spatial misjudgments in 2D-only laparoscopic training. We introduce a synchronized visualization framework that combines traditional 2D endoscopic views with real-time 3D visual feedback using NVIDIA Isaac Sim and the Meta Quest 3 MR headset. By integrating AI models for surgical tool localization~\cite{benavides2024real} and instrument-tissue interaction detection~\cite{lin2024instrument} within Isaac Sim’s Action Graph system, our work identifies spatial errors and provides instant corrective feedback using trajectory visualization, contact highlighting, and textual guidance. Our proposed framework supports the development of 3D spatial awareness without requiring specialized intraoperative hardware, bridging the gap between traditional training modalities and perceptually rich simulation environments.

\section{Methodology}
\subsection{Overview}

\begin{figure*}
    \centering    \includegraphics[width=1\linewidth]{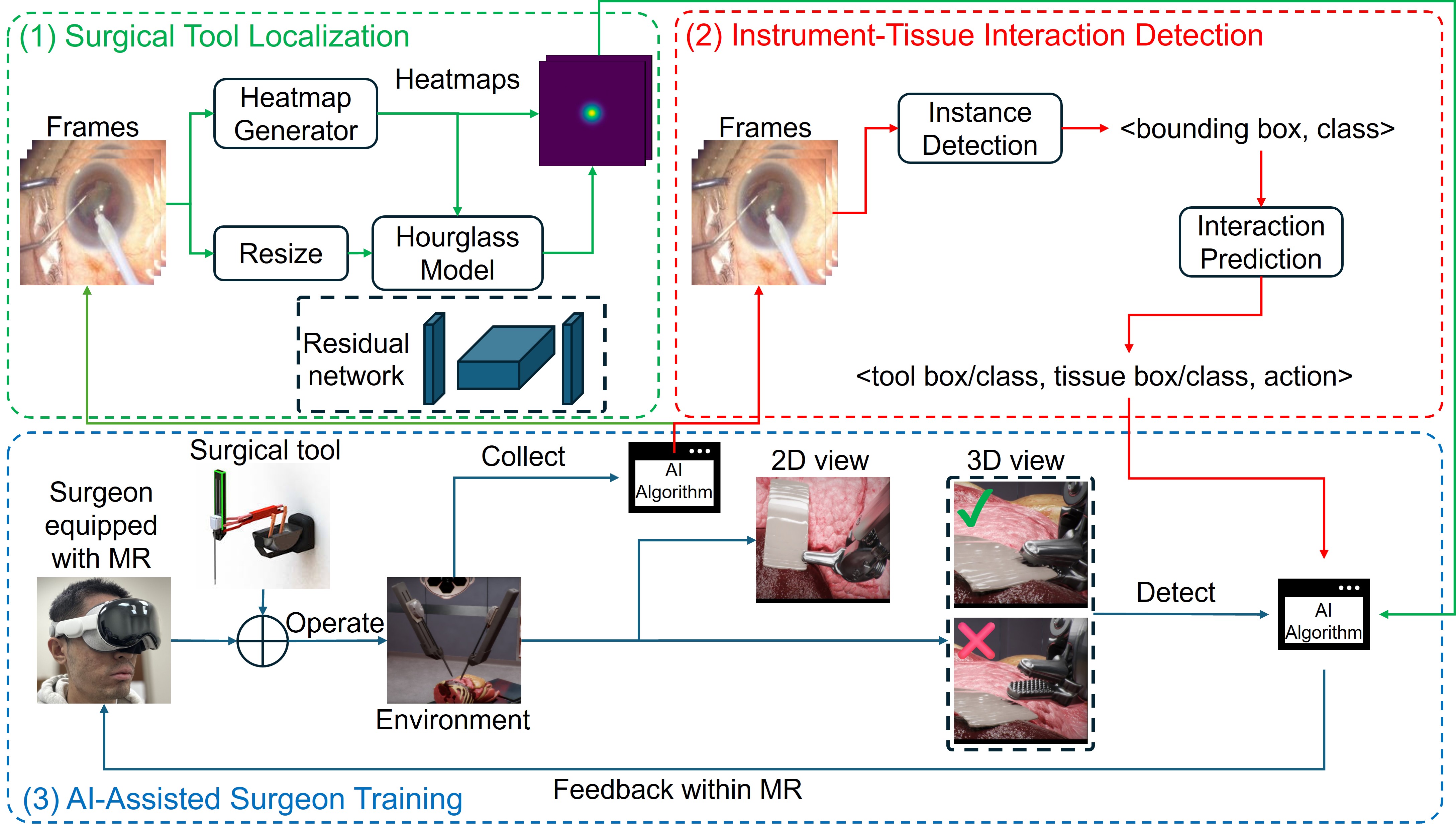}
    \caption{Overview of our proposed AI-assisted surgical training pipeline. The pipeline consists of three integrated modules. (1) Surgical Tool Localization (green box): Endoscopic video frames are resized and processed through an Hourglass convolutional network to generate heatmaps indicating tool tip locations. (2) Instrument-Tissue Interaction Detection (red box): A detection network identifies bounding boxes and classes of instruments and tissues, then predicts their interactions using temporal and spatial reasoning. (3) AI-Assisted Surgeon Training with Mixed Reality (blue box): A surgeon operates within a simulated environment while viewing standard 2D laparoscopic video. Real-time AI analysis provides 3D feedback via a mixed-reality headset, highlighting correct and incorrect tool-tissue interactions to improve spatial awareness and surgical precision.}
    \label{fig:2}
\end{figure*}

\begin{figure}[ht]
    \centering
\includegraphics[width=0.6\linewidth]{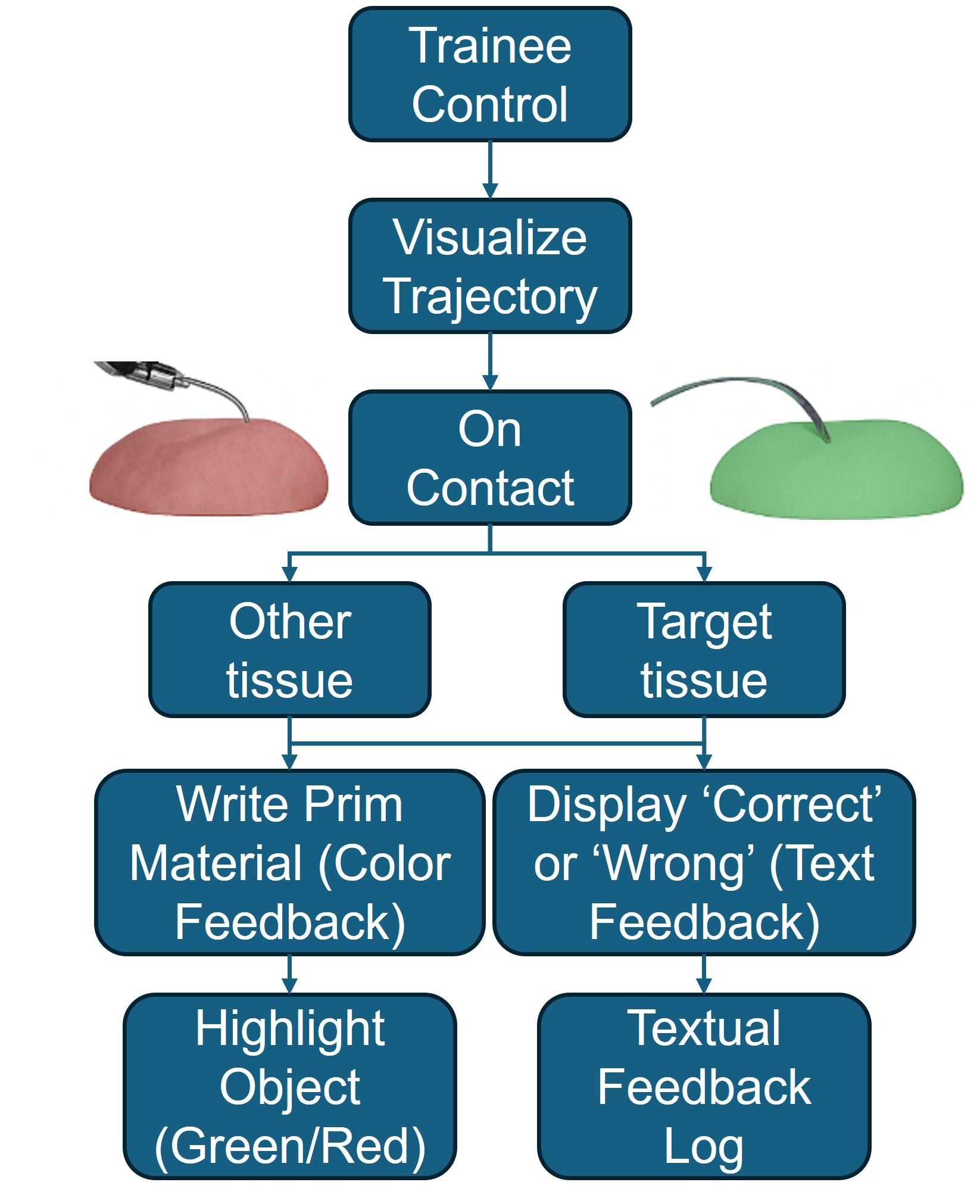}
    \caption{Action graph-based real-time visual feedback mechanism implemented in NVIDIA Isaac Sim. The trainee controls daVinci medical robotic arm (Trainee Control), which drives the end-effector toward the target while visualizing its trajectory. When a contact event is detected (On Contact), the system provides multiple visual feedback channels: color-coded material updates (Write Prim Material), screen-space text cues (Display ‘Correct’ or ‘Wrong’), and textual logging, to improve spatial awareness and guide correct surgical operations.}
    \label{fig:3}
\end{figure}

Fig.~\ref{fig:2} and Fig.~\ref{fig:3} illustrate the overall architecture of our proposed AI-assisted surgical training framework, designed to enhance spatial awareness during laparoscopic skill acquisition. The system consists of two main modules: an AI-based perception and analysis pipeline and an action graph–driven visual feedback mechanism implemented in the NVIDIA Isaac Sim environment. The two modules enable synchronized visualization, real-time tool-tissue interaction analysis, and MR feedback delivery through the Meta Quest 3 headset.

In the first stage (Fig.~\ref{fig:2}), the perception pipeline performs surgical tool localization and instrument-tissue interaction detection. Endoscopic image frames are processed through an hourglass-based residual network to generate heatmaps representing the estimated tool tip positions. Simultaneously, an instance detection module identifies the bounding boxes and classes of instruments and tissues, followed by temporal and spatial reasoning to infer their interaction states. The resulting metadata (\textit{tool class}, \textit{tissue class}, \textit{interaction type}) are transmitted to the Isaac Sim environment for visualization and feedback generation.

In the second stage (Fig.~\ref{fig:3}), these AI-derived spatial cues drive the real-time visual feedback pipeline built with the Action Graph system in Isaac Sim. The trainee controls a daVinci medical robotic arm to perform various surgical tasks. As the end-effector approaches the target, the system visualizes its trajectory using the Draw Debug Line node. Upon contact, On Contact Event triggers multiple visual feedback channels:
(1) color-coded material updates via Write Prim Material to mark correct (green) or incorrect (red) interactions,
(2) screen-space text overlays such as “Correct operation” or “Unsafe depth,” and
(3) optional textual logging for training analysis.

These cues are simultaneously displayed in the MR headset, allowing the trainee to perceive spatial relationships between the tool and surrounding anatomy in both 2D and 3D contexts. The tight integration between AI inference and physics-based simulation establishes a closed feedback loop that transforms conventional passive training into an interactive, perceptually guided learning experience.

\begin{figure*}
    \centering
\includegraphics[width=1\linewidth]{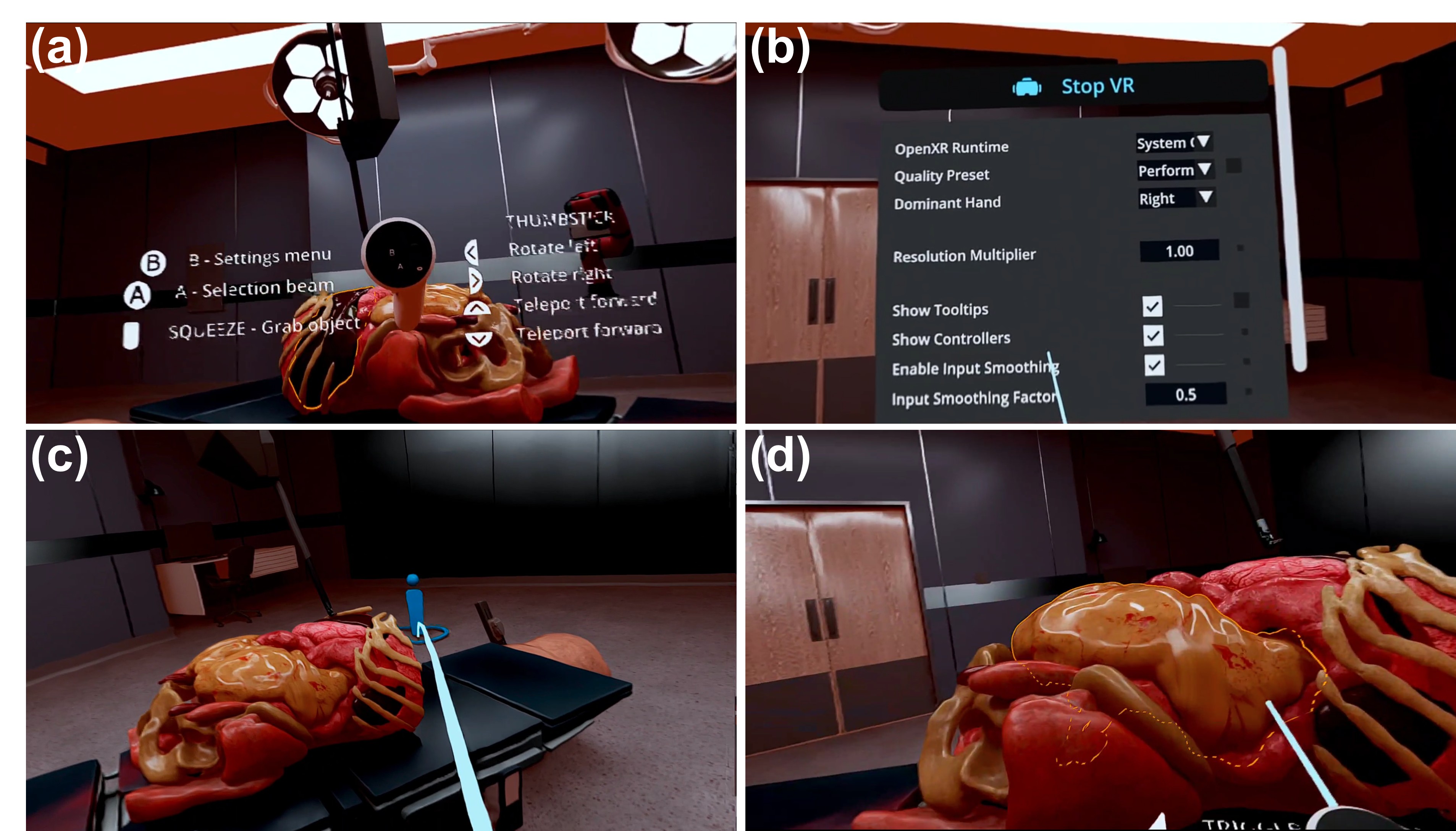}
    \caption{Trainee mixed-reality (MR) views during the laparoscopic training process implemented in NVIDIA Isaac Sim.
(a) Surgical room environment setup with controller options displayed.
(b) MR configuration interface for environment initialization and parameter adjustment.
(c) Trainee navigation within the MR scene using handheld controllers; the blue line visualizes the navigation trajectory.
(d) Object interaction and selection using controllers; the selected organ is highlighted with yellow dashed contours.}
    \label{fig:4}
    \vspace{-10pt}
\end{figure*}

\begin{figure}[htbp]
    \centering
\includegraphics[width=1\linewidth]{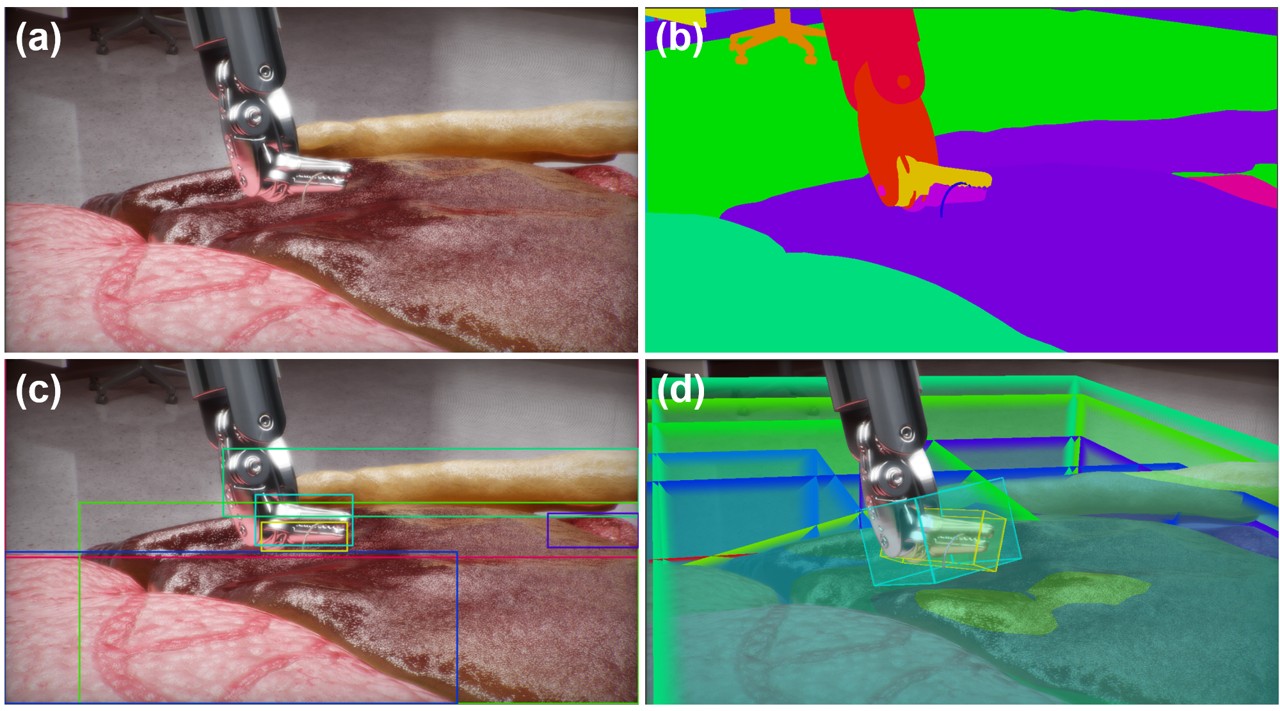}
    \caption{Multi-view visualization during surgical task. (a) Standard photorealistic rendering showing the daVinci robotic arm holding a needle. (b) Semantic segmentation view illustrating class-level pixel labeling for instruments and organs. (c) 2D bounding box visualization for key objects, including the robotic arm, needle, and anatomical structures. (d) 3D bounding box representation highlighting object geometry and spatial relationships within the simulation environment.}
    \label{fig:5}
    \vspace{-4pt}
\end{figure}

\begin{figure}[htbp]
    \centering
\includegraphics[width=1\linewidth]{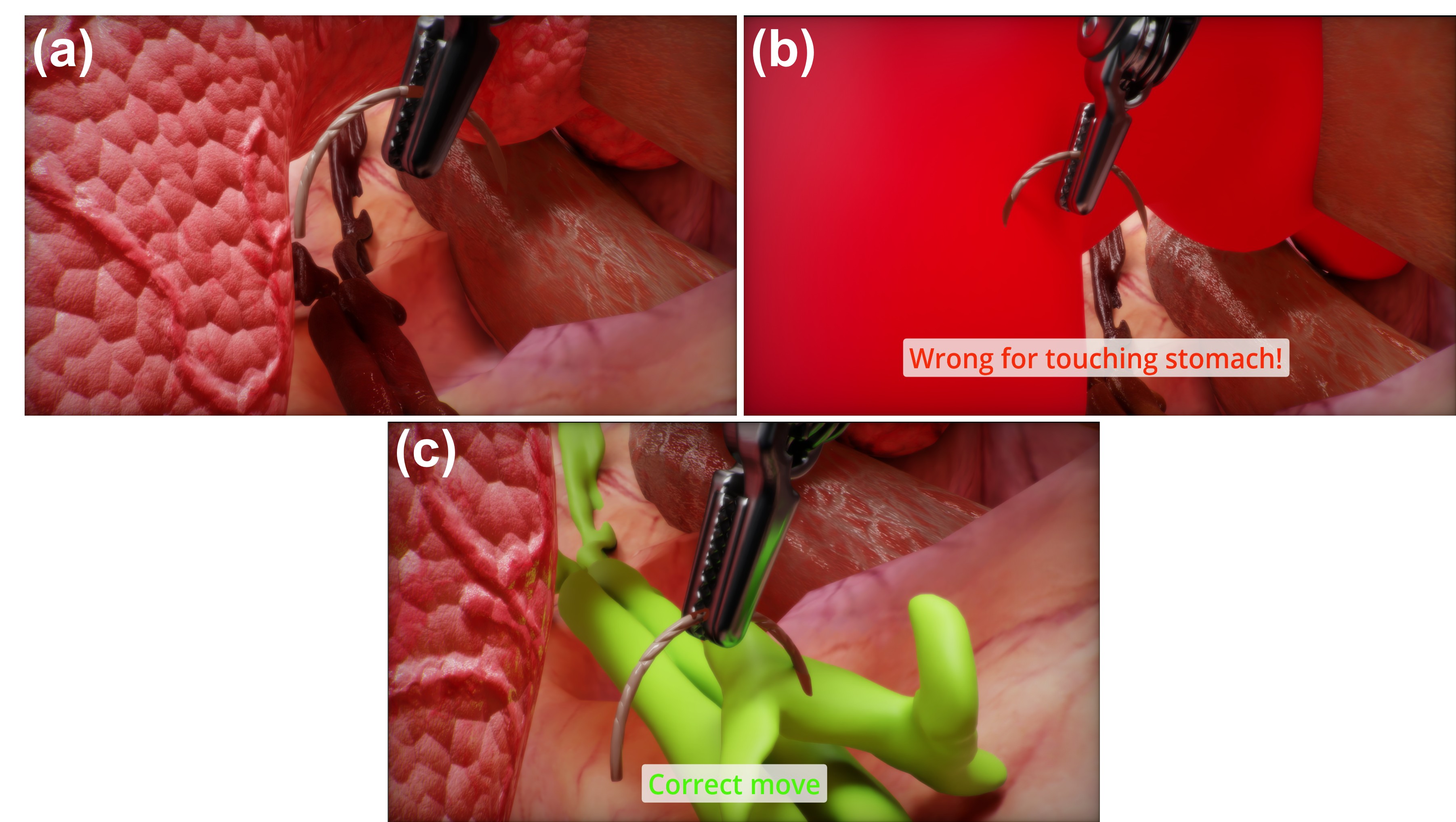}
    \caption{Example of real-time visual feedback during surgical task. (a) The robotic arm holding needle approaches the target tissue under normal operation. (b) When the needle contacts an incorrect object (e.g., the stomach), the system triggers a red overlay and displays an on-screen warning (“Wrong for touching stomach!”) using Action Graph nodes such as Write Prim Material and Draw Screen Text. (c) When the tool follows a correct trajectory and contacts the intended tissue, the object turns green and a “Correct move” message appears, providing intuitive visual feedback to reinforce spatial awareness and safe manipulation behavior.}
    \label{fig:6}
    \vspace{-40pt}
\end{figure}

\subsection{AI-Assisted Training Procedure}
Our framework integrates perception, simulation, and MR visualization into five structured stages that progressively develop 3D spatial awareness during laparoscopic training.

\textbf{Stage 1: Surgeon Setup and MR Environment Initialization.}
The trainee wears a Meta Quest 3 headset that enables rendering in NVIDIA Isaac Sim.
As shown in Fig.~\ref{fig:4}, the MR workspace is configured to replicate a realistic surgical room, including organ models, instruments, and control interfaces.
Trainees can interact with the virtual environment through handheld controllers—visualized navigation trajectories (Fig.~\ref{fig:4}c) and highlighted object selections (Fig.~\ref{fig:4}d) ensure intuitive spatial orientation before training begins.

\textbf{Stage 2: Multi-View Visualization and Scene Perception.}
During surgical tasks, the daVinci robotic arm and needle operate within the simulated anatomy.
Isaac Sim provides synchronized visualization channels for both trainee comprehension and AI processing, as illustrated in Fig.~\ref{fig:5}:
(a) the photorealistic endoscopic rendering; (b) the semantic-segmentation view labeling each instrument and tissue; (c) 2D bounding boxes for key objects; and (d) 3D bounding boxes that depict spatial geometry and relative depth.
These complementary representations establish the foundation for accurate spatial tracking and context-aware perception.

\textbf{Stage 3: Instrument-Tissue Interaction Detection.}
When the needle approaches or contacts tissue, Isaac Sim’s \textit{On Contact Event} node automatically records contact pairs, normals, and depths, allowing the system to determine whether the interaction is correct.
This process parallels AI-based frameworks such as ITID-net~\cite{lin2024instrument}, which formalize each interaction as a structured tuple including instrument type, tissue type, and contact region.
The captured data drive subsequent feedback generation in both the MR display and simulation log.

\textbf{Stage 4: Real-Time Visual Feedback through Action Graph.}
Upon detecting contact, Isaac Sim’s Action Graph dynamically triggers feedback nodes—\textit{Write Prim Material} to update object colors and \textit{Draw Screen Text} to display contextual messages.
As shown in Fig.~\ref{fig:6}, when the needle touches an incorrect structure (e.g., the stomach), a red overlay and warning (“Wrong for touching stomach!”) appear instantly.
On the contrast, correct tool-tissue interaction turns the contacted region green with a “Correct move” message.
These multimodal cues provide immediate, intuitive reinforcement of proper spatial alignment, depth control, and safe maneuvering.

\textbf{Stage 5: Iterative Practice and Skill Reinforcement.}
The trainee repeats the suturing sequence while observing MR feedback in real time.
Each interaction—color response, text message, or trajectory cue—is logged for post-training evaluation.
This iterative loop transforms passive observation into an interactive, feedback-driven learning process that strengthens spatial awareness, precision, and procedural confidence without requiring a physical trainer or patient model.

\subsection{Surgical Tool Localization}
To enable accurate real-time tracking of surgical instruments during laparoscopic procedures, we integrate the deep learning-based tool localization method~\cite{benavides2024real} into our training framework. This model builds upon a dual Hourglass network architecture, which preserves spatial information across scales through symmetric downsampling and upsampling pathways linked by skip connections. The network accepts normalized RGB endoscopic images, which are resized to fit the expected input shape. It then generates a 128×128 heatmap indicating the probable location of the tool center, with the highest pixel intensity corresponding to the predicted instrument tip. During training, corresponding ground-truth heatmaps are created by mapping the true tool center coordinates and refining them using a Gaussian-based spatial distribution tailored to the instrument’s size. The network contains multiple CNN branches, but in our application, only the tool-center prediction branch is retained for efficiency. Additionally, the original residual modules are replaced with fire modules to reduce computational cost by emphasizing channel-wise relationships before spatial filtering. This CNN-based model is used throughout our simulation to continuously localize surgical instruments from monocular endoscopic video in real time, providing a critical input for the AI system to assess whether the surgeon’s instrument positioning aligns with the intended anatomical targets.

\subsection{Instrument-Tissue Interaction Detection}  
To provide intelligent guidance during surgical training, our system integrates ITID-Net~\cite{lin2024instrument}, a dedicated deep learning framework for detecting fine-grained instrument-tissue interactions in laparoscopic video. Unlike prior works that represent interactions in a coarse and purely classificatory manner, ITID-Net explicitly models both spatial and semantic relationships between surgical tools and anatomical structures. It encodes each detected interaction as a quintuple—$\langle$instrument class, instrument bounding box, tissue class, tissue bounding box, action class$\rangle$—thereby offering detailed, localized understanding of surgical actions. ITID-Net includes several specialized components: (1) a Snippet Consecutive Feature (SCF) layer to enrich proposal-level representations using temporal context from adjacent video frames; (2) a Spatial Corresponding Attention (SCA) layer to align features across neighboring frames through spatial encoding; and (3) a Temporal Graph (TG) layer that builds intra-frame and inter-frame connections to capture both structural and temporal dependencies among instruments and tissues. Together, these modules enable real-time, context-aware detection of whether a surgical tool is appropriately interacting with the intended tissue. If the model identifies a mismatch, such as a tool acting upon the wrong tissue or failing to make effective contact, it triggers visual feedback within the MR environment, guiding the trainee to correct the spatial or functional misalignment.

\section{Spatial Awareness in Surgical Tasks}
Various simulated tasks have been proposed for a trainee to practice surgical skills effectively. They form the basis for skill acquisition~\cite{hong2021simulation}. In Oropesa et al.~\cite{oropesa2011methods}, a classification of basic tasks is presented. The existing simulation-based training systems use at least one of the five tasks listed below.

\subsection{Navigation}
Our AI assisted training uses the tool localization heatmaps and the interaction detector to teach depth and orientation during camera and instrument navigation while the trainee still views a standard 2D endoscopic feed. When the end effector drifts or the camera pose creates occlusion the system infers the 3D relation between the tool tip and landmarks from the heatmaps and the detected interaction tuple and then presents mixed reality cues such as a safe viewing cone, distance bars, and a suggested approach vector. By repeatedly mapping 2D visual cues to these 3D prompts the trainee learns to calibrate depth, compensate for the fulcrum effect, and maintain stable spatial reference frames.

\subsection{Object manipulation}
For grasp and pull tasks the AI continuously checks whether the jaws contact the intended target and whether traction is applied along a safe 3D direction. If contact is missed or the pull is misdirected the headset overlays a grasp point marker and a 3D arrow that indicates the correct traction vector while the 2D video shows a soft highlight at the missed contact. This coupling of detection and guidance helps trainees internalize where the tool is in depth relative to the tissue and how to align force along a desirable spatial axis.

\subsection{Object transfer}
During peg transfer and needle passing the framework tracks both instruments and the object and uses the interaction detector to verify clean release and acquisition. When the receiving tool is offset in depth or orientation the headset renders a ghost target pose and a 3D handoff corridor that the user should enter, while the 2D feed remains unchanged to preserve clinical realism. This teaches bimanual spatial coordination, distance estimation between tools, and precise alignment during approach and handoff.

\subsection{Cutting related tasks}
For cutting and dissection, our AI assisted pipeline uses tool localization heatmaps together with instrument tissue interaction detection to verify that the scissor blades are in contact with the intended tissue layer and to compare the observed blade path against a planned surface trajectory. When the trainee cuts air or drifts above or below the target plane, the mixed reality headset renders a transparent three dimensional cutting plane and an on path corridor, while the two dimensional endoscopic view receives minimal off target annotations. By repeatedly pairing two dimensional visual cues with explicit three dimensional references, the trainee learns to recognize tissue planes in depth, to set an appropriate entry depth, and to maintain adherence to the planned path.

\begin{figure}
    \centering
    \includegraphics[width=1\linewidth]{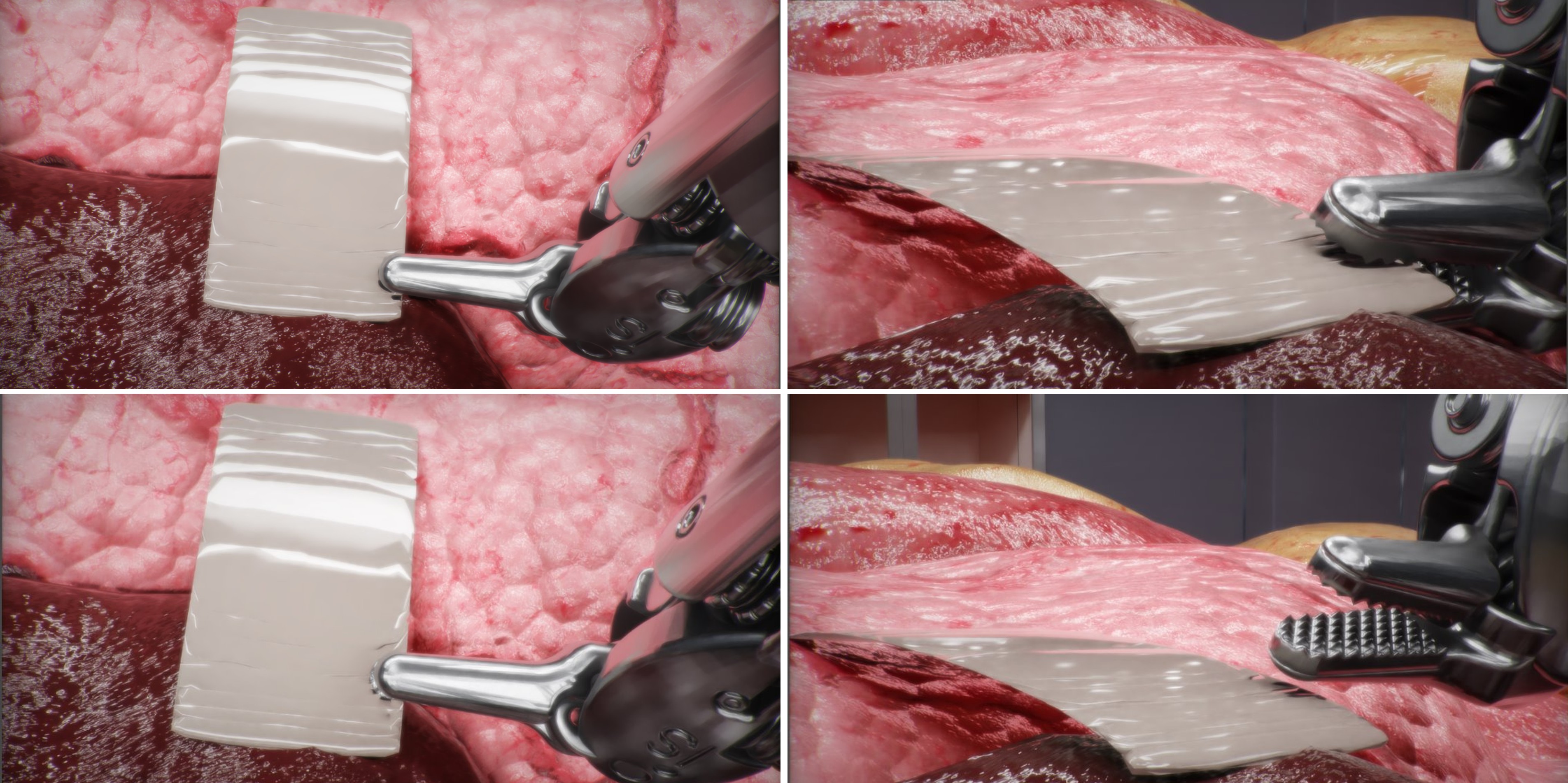}
    \caption{Cutting task with synchronized two dimensional and three dimensional views. Left column: the two dimensional endoscopic view shown to the surgeon during operation or training. Right column: the synchronized three dimensional scene. Top row: correct execution with the tool in contact with the target tissue and aligned with the cutting plane. Bottom row: incorrect execution where the tool cuts air and does not contact the target. The two dimensional views in the two rows look nearly identical, which makes depth judgment difficult. By adding the three dimensional context, the proposed method reveals the true tool target relationship and helps the surgeon learn spatial awareness.}
    \label{fig:7}
\end{figure}

Within NVIDIA Isaac Sim we recreate the clipping and cutting of the cystic artery during laparoscopic cholecystectomy to study these effects. In a traditional monocular endoscopic view, key anatomical landmarks can be obscured and depth estimation is difficult. In contrast, our pipeline presents synchronized real time three dimensional spatial visualization alongside the conventional two dimensional laparoscopic video, revealing the relative positions of instruments and targets and directly resolving depth ambiguity. Preliminary visual analysis of these simulated scenes indicates that the combined two dimensional and three dimensional presentation reduces spatial misjudgment by helping trainees map planar images to stable three dimensional geometry.
Fig.~\ref{fig:7} shows the cutting task with synchronized two dimensional and three dimensional views. The left panels show the two dimensional endoscopic view presented to the trainee, while the right panels show the synchronized three dimensional scene. Although the two dimensional views look nearly identical for the correct (top) and incorrect (bottom) trials, the three dimensional context reveals whether the tool truly contacts the target or is cutting air, demonstrating how our method strengthens spatial awareness.

\subsection{Suturing related tasks}
Suturing requires precise spatial orientation and careful instrument manipulation, yet traditional two dimensional endoscopic views provide limited depth cues and often lead to errors in needle entry depth and incident angle. In our AI assisted pipeline, the tool localization module delivers stable estimates of the needle tip and pose, while the instrument tissue interaction detector verifies valid tissue engagement at both entry and exit. When the angle or depth is incorrect, the mixed reality headset projects three dimensional entry and exit markers and a curved trajectory consistent with the needle radius, while the two dimensional view remains the operative reference. This pairing of minimal clinical visuals with explicit three dimensional guidance teaches trainees to judge tissue thickness, choose an appropriate incident angle, and follow a spatial arc that achieves reliable tissue approximation without excessive force.

\begin{figure}
    \centering
    \includegraphics[width=1\linewidth]{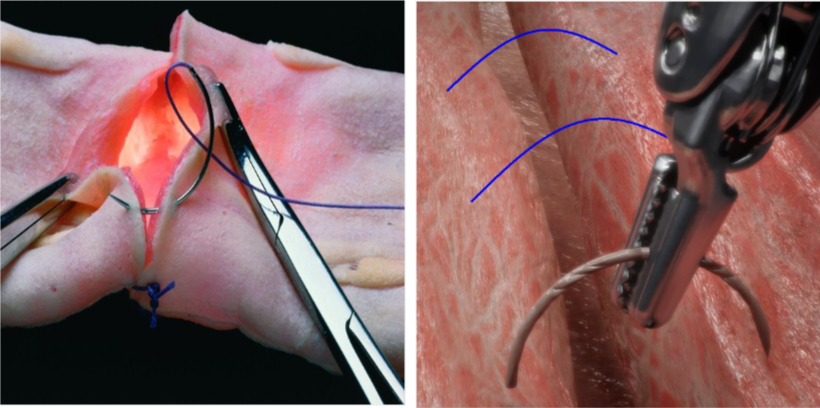}
    \caption{Suture-related training in reality (Left) and simulation (Right). Left: real-world bench task for needle handling and tissue approximation~\cite{suture}. Right: our Isaac Sim scene, where the blue curves indicate the desired needle driving arc and tool trajectory for correct entry, depth, and exit, providing explicit three dimensional guidance for spatial awareness learning.}
    \label{fig:8}
\end{figure}

Within Isaac Sim we construct a realistic laparoscopic suturing scene to evaluate this training strategy. When relying only on the conventional two dimensional visualization, trainees frequently place the needle too shallow or too deep and approach at suboptimal angles. With the synchronized two dimensional and three dimensional presentation, the system provides immediate feedback on needle orientation, insertion depth, and the spatial relationships among the needle, suture, and tissue. These cues reduce cognitive effort otherwise spent on mentally reconstructing geometry and enable more accurate and consistent needle placement, leading to improved suturing precision and efficiency. Fig.~\ref{fig:8} shows the suture related training in reality and simulation. The left panel shows a real world bench task for suturing, with needle handling and tissue approximation. The right panel is our Isaac Sim scene where blue curves indicate the desired needle driving arc and tool trajectory, providing explicit three dimensional guidance that supports spatial awareness training.

\section{Conclusion and Future Work}
In this work, we present an AI assisted training framework that improves spatial awareness for laparoscopic surgery by coupling a conventional two dimensional endoscopic view with synchronized three dimensional cues delivered through mixed reality. While the trainee operates using only the clinical two dimensional feed, AI modules run in the background to localize tools and to detect instrument tissue interactions, allowing the system to infer whether contact, distance, and orientation are correct for the current task. When a spatial misjudgment is likely, the headset renders concise three dimensional guidance such as safe viewing cones, approach corridors, entry and exit markers, and curved needle paths, while the two dimensional view remains the operative reference. The same mechanism applies across key tasks including navigation, object manipulation, object transfer, cutting, and suturing. Simulation results in Isaac Sim show that scenes which appear identical in two dimensions can be disambiguated by the added three dimensional context, supporting learning of depth, handling of occlusion, compensation for the fulcrum effect, and adherence to planned trajectories.

While our work demonstrates potential for enhancing spatial awareness and training efficiency, several limitations remain. Our system is developed solely within a controlled simulation environment, and further work is needed to evaluate its deployment feasibility across different surgical organizations and training institutions. Practical adoption may be limited by hardware availability, computational requirements, and the need for technical expertise to integrate the Isaac Sim-based pipeline into existing curricula.

For future research, several promising directions can be pursued to expand and validate this pipeline. First, real-time patient motion tracking could be integrated to further enhance simulation realism, ensuring the safety and optimal positioning of patients by dynamically adjusting visualizations according to patient movements. Second, the systematic recording and analysis of detailed surgical techniques performed within this dual-view framework could provide valuable data-driven insights for educational feedback, potentially accelerating skill development and standardizing surgical practices. Finally, incorporating precise surgical instrument tracking algorithms into the pipeline could allow continuous analysis of instrument trajectory and location, significantly increasing the accuracy, realism, and training efficacy of the system. These advancements together could significantly enhance the realism, effectiveness, and clinical applicability of our proposed spatial training platform.

\bibliographystyle{IEEEtran}
\bibliography{export}
\end{document}